\documentstyle[11pt,epsf]{article}

\def\mypagenumber{1}
\def\mydate{February 5, 1996}
\def\myend{\end{document}}

\normalsize

\newcounter{sxn}

\newcounter{axn}

\date{}

\newdimen\mybaselineskip
\newcommand{\beeq}{\begin{equation}}
\newcommand{\eneq}{\end{equation}}
\newcommand{\beqn}{\begin{eqnarray}}
\newcommand{\eeqn}{\end{eqnarray}}

\def\mybig{\displaystyle \strut }

\def\dd{\partial}
\def\la{\raise.16ex\hbox{$\langle$} \, }
\def\ra{\, \raise.16ex\hbox{$\rangle$} }
\def\ran{\raise.16ex\hbox{$\rangle$} }
\def\go{\rightarrow}

\def\next{{~~~,~~~}}
\def\onehalf{ \hbox{${1\over 2}$} }

\def\psibar{ \psi \kern-.65em\raise.6em\hbox{$-$}\lower.6em\hbox{} }
\def\mbar{ m \kern-.78em\raise.4em\hbox{$-$}\lower.4em\hbox{} }

\def\Bbar{ B \kern-.73em\raise.6em\hbox{$-$}\hbox{} }

\def\L{ {\cal L} }
\def\ep{\epsilon}

\def\wil{ \Theta_{\rm W} }
\def\Pw{ P_{\rm W} }
\def\vphi{ {\varphi} }

\def\tchi{{\tilde \chi}}

\def\EM{{\rm EM}}
\def\tot{{\rm tot}}
\def\eff{{\rm eff}}
\def\mass{{\rm mass}}

\def\vac{{\rm vac}}
\def\rint{{\rm int}}
\def\ext{{\rm ext}}

\def\in{{\rm in}}
\def\out{{\rm out}}

\def\LapN{{\triangle_N^\vphi}}
\def\potN{{V_N(\vphi;\theta_\eff)}}

\def\myfrac#1#2{{\mybig #1\over \mybig #2}}

\def\boxit#1{$\vcenter{\hrule\hbox{\vrule\kern3pt
     \vbox{\kern3pt\hbox{#1}\kern3pt}\kern3pt\vrule}\hrule}$}
\def\bigbox#1{$\vcenter{\hrule\hbox{\vrule\kern5pt
     \vbox{\kern5pt\hbox{#1}\kern5pt}\kern5pt\vrule}\hrule}$}

\begin{document}

\bibliographystyle{unsrt}
\footskip 1.0cm
\thispagestyle{empty}
\setcounter{page}{\mypagenumber}

{\baselineskip=10pt \parindent=0pt \small
\mydate 
\hfill \hbox{\vtop{\hsize=3.5cm  hep-th/
    \\ UMN-TH-1421/96\\  NUC-MINN-96/2-T \\}}

\vspace{15mm}
}

\centerline {\Large\bf  Confinement and Chiral Condensates in 2-d QED}
\vspace*{4mm}
\centerline {\Large\bf  with Massive N-flavor Fermions}

\vspace*{10mm}
\centerline {\large  Ram\'on Rodr\'\i guez$^{1}$  and Yutaka Hosotani$^{2}$}

\vspace*{5mm}
\centerline {\small\it School of Physics and Astronomy, University
       of  Minnesota}
\centerline {\small\it Minneapolis, Minnesota 55455, U.S.A.}
\vspace*{3mm}
\centerline {\small\sf $^1$rodriguez@physics.spa.umn.edu}
\centerline {\small\sf $^2$yutaka@mnhep.hep.umn.edu}
\bigskip
\vspace*{5mm}
%\baselinestretch{2.0}
\normalsize

\begin{abstract}
\baselineskip=13pt
We evaluate Polyakov loops and string tension in two-dimensional 
QED with both massless and massive  $N$-flavor fermions at 
zero and  finite temperature.  External charges, or external electric fields, 
induce phases in fermion masses and shift the value of
the vacuum angle parameter $\theta$, which in turn alters the chiral
condensate.   In particular, in the presence of two sources of opposite
charges, $q$ and $-q$,  the shift in $\theta$ is $2\pi(q/e)$ 
independent of $N$.  The string tension has a cusp singularity at $\theta=\pm\pi$
for $N\ge 2$ and is proportional to
$m^{2N/(N+1)}$  at $T=0$.
\end{abstract}

%\end{titlepage}
 
\vspace*{10mm}

%\newpage

%\setcounter{page}{1}

%\textheight=20cm
%\headsep=0.75cm
%\vsize=20cm
\baselineskip=16pt

%%%%%%%%%%%%%%%%%%%%%%%%%%%%%%%%%%%%%%%%%%%%%%%%%%%%%%%%%%%%%%%%%

Two-dimensional QED, the Schwinger model,  with massive $N$-flavor fermions
resembles four-dimensional QCD in various aspects, including confinement, chiral
condensates, and $\theta$ vacua \cite{SchwGEN}-\cite{Gattringer1}. 
Much progress has been made recently
in evaluating chiral condensates and string tension in the massive
theory \cite{HNZ}-\cite{Smilga2}. 
In this paper we shall show that the three phenomena,  confinement, chiral
condensates, and $\theta$ vacua, are intimately related to each other.
In particular, the string tension in the confining potential is determined by  
the $\theta$ dependence of chiral condensates $\la \psibar\psi\ra$.

The behavior of the model is distinctively different, depending on
whether $N=1$ (one-flavor) or $N\ge 2$ (multi-flavor), and on whether
fermions are massless or massive.    The massless ($m=0$) theory 
is exactly solvable.   $\la \psibar\psi\ra_\theta \not= 0$
for $N=1$, but $\la \psibar\psi\ra_\theta = 0$ for $N\ge 2$.  
\cite{Halpern,Affleck}    In either
cases the string tension between two external sources of opposite
charge vanishes  \cite{CJS,Ellis,HNZ}.

In the massive ($m\not= 0$) theory 
$\la \psibar\psi\ra_\theta$ is proportional to 
$ m^{(N-1)/(N+1)} \cos^{2N/(N+1)} (\bar\theta/N)$ at $T=0$ where 
$\bar \theta=\theta- 2\pi[(\theta+\pi)/2\pi]$ \cite{Coleman,HHI}. 
For $N\ge 2$ the dependence
on $m$ is non-analytic. It also has a cusp singularity at $\theta=\pm\pi$.
A perturbation theory in fermion masses is not  valid at low temperature.

The confinement phenomenon can be explored in various ways.  One way is to
evaluate the Polyakov loop $P_q(x) = \exp \big\{ iq \int_0^\beta d\tau \, 
A_0 (\tau,x) \big\}$  at finite temperature $T=\beta^{-1}$ 
\cite{Polyakov}-\cite{Grignani2}.
$F(T) = -T \ln \la P_q(x) \ra$ or $ -T \ln \la P_q(x)^\dagger P_q(y) \ra$
measures the increase in free energy in the presence of an external
charge $q$ or a pair of charges $q$ and $-q$.  In particular, the latter
is written as $\sigma |x-y|$ for large $|x-y|$ where $\sigma$ is 
identified with the string tension.  This method has the  advantage of 
giving the temperature dependence directly.

Alternatively one may determine the ground state to evaluate the change in the
energy density (at $T=0$) when a pair of sources of charge $q$ and $-q$ is
placed \cite{CJS,HH,Iso}.   This method has the
advantage of showing how external charges affect the $\theta$ parameter
and chiral condensates.

We employ both methods  in a unified manner.  Years ago Coleman,
Jackiw, and Susskind showed the confinement of fractional charges
in the $N=1$ theory adopting the latter method \cite{CJS}. 
Recently Hansson,
Nielsen and Zahed applied functional integration method to evaluate the 
Polyakov loop correlation function \cite{HNZ}.  
The argument has been generalized to
finite temperature by Grignani et al. \cite{Grignani1}.  
Ellis et al. \cite{Ellis} and Gross et al. \cite{Gross} 
have presented the mechanism of confinement in terms of soliton
solutions in the bosonized form.  All of these arguments are given in
the one flavor ($N=1$) case and rely on the validity of a perturbation
theory in a fermion mass.  

Recently   chiral condensates with arbitrary  fermion masses $m$,
 vacuum angle $\theta$ and temperature $T$ have been evaluated in the
$N$-flavor model \cite{HHI}.  The problem was reduced to solving a quantum
mechanical system of $N$ degrees of freedom.   It was shown that the $m\go 0$
and $T\go 0$ limits do not commute for  $N\ge 2$.  In particular, the
$m$-dependence of  physical quantities is singular at $T=0$.

We analyse the model
\beqn
&&{\cal L} = - \hbox{$1\over 4$} \, F_{\mu\nu} F^{\mu\nu} + 
\sum_{a=1}^N \psibar_a \Big\{ \gamma^\mu (i \dd_\mu - e A_\mu)  \Big\} 
   \psi_a + {\cal L}_\mass \cr
&&{\cal L}_\mass =  - \sum_{a=1}^N m_a 
\Big\{ e^{i\delta_a} \, M_a +  e^{-i\delta_a} \,M_a^\dagger \Big\} 
   \hskip 1cm (m_a \ge 0)\cr
&&M_a = \psibar_a \onehalf(1-\gamma^5)\psi_a  
  \label{Lagrangian}
\eeqn
defined on a circle  with a circumference $L$
\cite{HH}-\cite{Barut}.   The model defined at
finite temperature \cite{Love}-\cite{Steele},  on a torus or
sphere \cite{Joos1}-\cite{Ferrari}, or on a lattice or light-cone
\cite{Lenz}-\cite{Hallin},  has been also extensively discussed in the
literature.

We impose boundary conditions $A_\mu(t,x+L)=A_\mu(t,x)$ and
$\psi_a(t,x+L) = - \psi_a(t,x)$.  On a circle the only physical degree
of freedom associated with gauge fields is the Wilson line phase 
$\wil(t)$: \cite{HH}
\beeq
e^{i\wil(t)} = \exp \bigg\{ ie\int_0^L dx \, A_1(t,x) \bigg\} ~~.
\label{Wilson1}
\eneq

In Matusbara's formalism the finite temperature field theory is defined 
by imposing perodic or anti-periodic boundary conditions in the imaginary
time ($\tau$) axis on bosons or fermions, respectively.  
Mathematically  the model at finite temperature $T=\beta^{-1}$ is
obtained from the model defined on a circle by Wick rotation and replacement
$L \go \beta$, $it \go x$ and $x\go \tau$.
Furthermore the Polyakov loop of a charge $q$ in the finite temperature 
theory corresponds to the Wilson line phase:
\beeq
P_q(x) = \exp \bigg\{ iq\int_0^\beta d\tau \, A_0(\tau, x) \bigg\}
\Longleftrightarrow \exp \bigg\{ i\, {q\over e} \, \wil(t) \bigg\}~~.
\label{PWcorrespondence}
\eneq

We  bosonize fermions  in the  Coulomb gauge
\cite{Coleman,Halpern,HH,HHI}.  Take
$\gamma^\mu = (\sigma_1, i\sigma_2)$ and write $\psi_a^T=(\psi^a_+,\psi^a_-)$.
In the interaction picture defined by free massless fermions 
\beqn
\psi^a_\pm (t,x) = {1\over \sqrt{L}} \, C^a_\pm \,
 e^{\pm i \{ q^a_\pm + 2\pi p^a_\pm (t \pm x)/L \} }
  :\, e^{\pm i\sqrt{4\pi} \phi^a_\pm (t,x) } \, :  &&\cr
\noalign{\kern 10pt}
e^{2\pi i p^a_\pm} ~ | \, {\rm phys} \ra = |\, {\rm phys} \ra
   \hskip 3cm &&
\label{bosonize}
\eeqn
where $C^a_+ =\exp \{ i\pi \sum_{b=1}^{a-1} ( p^b_+ + p^b_-)\}$ and 
$C^a_- = \exp \{ i\pi \sum_{b=1}^{a} ( p^b_+ - p^b_-) \}$.  Here
\beqn
[q^a_\pm, p^b_\pm] &=& i \, \delta^{ab}\cr
\phi^a_\pm (t,x) &=& \sum_{n=1}^\infty (4\pi n)^{-1/2} \,
  \big\{ c^a_{\pm,n} \, e^{- 2\pi in(t \pm x)/L} + {\rm h.c.} \big\}\next
[c^a_{\pm,n}, c^{b,\dagger}_{\pm,m}] = \delta^{ab} \delta_{nm}~~~.
\label{CR}
\eeqn
The $:~:$ in  (\ref{bosonize})  indicates normal ordering with respect to
$(c_n^{},c_n^\dagger)$.    In physical states $p^a_\pm$ takes an integer
eigenvalue.

Conjugate pairs are
$\{p_a,q_a\}=\big\{ \onehalf (p^a_+ + p^a_-),  q^a_+ + q^a_-  \big\}$, 
$\{\tilde p_a, \tilde q_a\} =\big\{ p^a_+ - p^a_-,\onehalf (q^a_+ -
q^a_-) \big\}$, $\{\Pw, \wil\}$,
and $\{ \Pi_a, \phi_a=\phi^a_++\phi^a_- \}$. The Hamiltonian in the
Schr\"odinger picture becomes
\beqn
&&H_\tot= H_0 + H_\phi  + H_\mass + ({\rm constant}) \cr
\noalign{\kern 8pt} 
&&H_0~  =  {e^2 L\over 2} \Pw^2 
 + {N\over 2\pi L} \Big\{ \wil + 
    {2\pi\over N} \sum_{a=1}^N p_a \Big\}^2  
%\cr &&\hskip 1cm 
- {2\pi\over NL} \Big\{ \sum_{a=1}^N p_a \Big\}^2
+ {2\pi\over L} \sum_{a=1}^N \Big\{ p_a^2 
   + {1\over 4} \tilde p_a^2 \Big\}\cr
\noalign{\kern 8pt} 
&&H_\phi = \int_0^L dx \, {1\over 2} :\,  \bigg[ ~ 
  \sum_{a=1}^N \Big\{ \, \Pi_a^2 + (\phi_a')^2 \Big\}
+{e^2\over \pi}  \Big( \sum_{a=1}^N \phi_a  \Big)^2 ~ \bigg] :\,  \cr
\noalign{\kern 8pt}
&&H_\mass = \int_0^L dx \, \sum_a m_a \Big\{ 
 e^{i\delta_a} M_a +  e^{-i\delta_a} M_a^\dagger \Big\} ~~~.
    \label{Hamiltonian}
\eeqn
In the mass term  
\beeq
M_a =
  - C^{a \dagger}_- C^a_+\cdot e^{ - 2\pi i \tilde p_a x/L}
\, e^{i q_a}  \cdot 
 L^{-1}  N_0[e^{i \sqrt{4\pi} \phi_a}]   
\label{massOperaor}
\eneq
where $N_\mu [\cdots]$ indicates that the operator inside $[ ~~]$ is
normal-ordered with respect to a mass $\mu$.  In general a mass-eigenstate
field $\chi_\alpha$ with a mass $\mu_\alpha$ is related to $\phi_a$ by an
orthogonal transformation $\chi_\alpha = U_{\alpha a} \phi_a$.
In (\ref{massOperaor}) we have
\beqn
&&N_0[e^{i\sqrt{4\pi}\phi_a}] = \Bbar_a \prod_{\alpha=1}^N 
  N_{\mu_\alpha} [ e^{iU_{\alpha a}\sqrt{4\pi}\chi_\alpha} ]  \cr
&&\Bbar_a = \prod_{\alpha=1}^N  B(\mu_\alpha L)^{{(U_{\alpha a})}^2} \cr
&&B(z) =
{z\over 4\pi} \exp \bigg\{ \gamma + {\pi\over z}
 - 2 \int_1^\infty  {du \over (e^{uz} - 1)\sqrt{u^2-1}}  \bigg\}~. 
\label{massOperator2}
\eeqn
As $[\tilde p_a, H_\tot] =0$, we may restrict ourselves to states with
$\tilde p_a=0$.

$H_\mass$ gives rather complicated coupling between the 
zero and $\phi_a$ ($\chi_\alpha$) modes,  whose effects are  
non-perturbative for $N\ge 2$.  As in previous papers \cite{HHI}
the vacuum wave function is written in the form
\beeq
|\Phi_\vac(\theta)\ra = \int_0^{2\pi} \prod_{a=1}^{N-1} d\vphi_a ~
   f(\vphi_a;\theta_\eff) ~ |\Phi_0(\vphi_a+\delta_a;\theta)\ra 
\label{vacuum1}
\eneq
where
\beqn
&&|\Phi_0(\vphi_a;\theta)\ra = (2\pi)^{-N/2} \sum_{ \{ n,r_a\} }
e^{in\theta + i\sum_{a=1}^{N-1}r_a\vphi_a } ~
 |\Phi_0^{(n+r_1, \cdots, n+r_{N-1},n)}\ra  \cr
&&\la\wil,q_a |\Phi_0^{(n_1, \cdots, n_N)}\ra =
 u_0\Big( \wil +  {2\pi\over N} \sum_{a=1}^{N} n_a \Big)
~(2\pi)^{-N/2} e^{i \sum_{a=1}^{N} n_a q_a } \cr
&&u_0(x) = \Big( {N\over \pi^2\mu L} \Big)^{1/4} e^{-N x^2/2\pi\mu L}    \next
\mu^2 = {Ne^2\over \pi}  \cr 
&&\theta_\eff = \theta - \sum_{a=1}^N \delta_a ~~~.
\label{vacuum2}
\eeqn 
We have generalized the expression to incorporate the phases $\delta_a$'s
in the mass parameter in (\ref{Lagrangian}).  The ansatz above for the vacuum
is good for $m_a \ll e$.
%We shall see below that the effect
%of external electric field  is absorbed in the redefinition of $\delta_a$.

The eigenvalue equation 
$(H_0+H_\mass) |\Phi_\vac(\theta) \ra=E |\Phi_\vac(\theta) \ra$  reads
\beeq
\Big\{ -\LapN + \potN \Big\} ~ f(\vphi;\theta_\eff)
= \ep ~ f(\vphi ; \theta_\eff) \label{QMeq}
\eneq
where
\beqn
\LapN &=& 
\sum_{a=1}^{N-1} {\dd^2\over \dd\vphi_a^2} 
-{2\over N-1} \sum_{a<b}^{N-1} {\dd^2\over \dd\vphi_a \dd\vphi_b} \cr
\noalign{\kern 5pt}
\potN &=& -{NL\over (N-1)\pi} \, e^{-\pi/N\mu L} 
~  \sum_{a=1}^N  m_a \Bbar_a  \cos \vphi_a  \cr 
\vphi_N &=&  \theta_\eff - \sum_{a=1}^{N-1} \vphi_a    
\label{LapPotN}
\eeqn
and $\ep = NEL/2\pi(N-1)$.

 For the $\chi_\alpha$ fields, the vacuum is defined with
respect to their physical mass $\mu_\alpha$'s which needs to be determined
self-consistently from the wave function in (\ref{vacuum1}).  In the symmetric
case $m_a = m$ one has $\mu_2=\cdots=\mu_N$, $\mu_1^2=\mu^2 +
\mu_2^2$ and $\Bbar_a = B(\mu_1 L)^{1/N} B(\mu_2 L)^{(N-1)/N} \equiv
\Bbar$.  The potential is reduced to
\beqn
&&\potN = - \kappa_0 \sum_{a=1}^N \cos \vphi_a \cr
&&\kappa_0 =  {Nm L\over (N-1)\pi} \, e^{-\pi/N\mu L} \, \Bbar~~. 
\label{Potential2}
\eeqn
Further $\mu_2^2$ is determined by
\beeq
\mu_2^2 = R = {8\pi m \Bbar\over L} \,  e^{-\pi/N\mu L} \, 
    \la \cos \vphi \ra_f 
   \label{bosonMass2}
\eneq
where the $f$-average is given by 
$\la g(\vphi) \ra_f = \int [d\vphi] \, g(\vphi)  |f(\vphi) |^2 $.   We have
made use of the fact that $\la e^{i\vphi_a} \ra_f$ is independent of $a$.
Eqs.\ (\ref{QMeq}) and (\ref{bosonMass2}) are solved simultaneously.
Evaluation of these equations was given in \cite{HHI}.  An important
point in the following discussion is that $f(\vphi;\theta_\eff)$ or
$\Phi_\vac(\theta)$ is determined  solely by $m$, $L$ (or $T$), and
$\theta_\eff$.

Now let us evaluate the Polyakov loop at finite $T$, or equivalently
$\la e^{ik\wil(t)}\ra_\theta$ on a circle. The parameter $k$ corresponds 
to $q/e$ where $q$ is the charge of an external source.  
Since the expectation value is $t$-translation invariant, it is sufficient 
to evaluate at $t=0$.  Making use of (\ref{vacuum1}), one immediately 
finds
\beeq
\la \Phi(\theta';\vphi')|e^{ik\wil}|\Phi(\theta;\vphi)\ra 
= \delta_{2\pi}(\theta-\theta'-2\pi k) 
\prod_{a=1}^{N-1}  \delta_{2\pi}(\vphi_a-\vphi_a'-{2\pi k\over N}) 
\cdot e^{-\pi k^2\mu L/4N}   ~~.
\label{Wilson2}
\eneq
It follows that
\beqn
&&\la P_{ke} \ra_{\theta,T} = \la e^{ik\wil}\ra_{\theta,L=T^{-1}} \cr
\noalign{\kern 12pt}
&&=
\cases{0 &for $k\not=$ an integer\cr
e^{-k^2\pi \mu /4NT} \mybig \int [d\vphi] \,
f(\vphi_a;\theta_\eff)^* 
  f(\vphi_a+{\mybig 2\pi k\over \mybig N}; \theta_\eff)
&for $k=$ an integer~~.\cr}
\label{Polyakov2}
\eeqn

The vanishing of the Polyakov loop for a fractional $k$ is  due to the
invariance  under large gauge transformations.  The Hamiltonian 
(\ref{Hamiltonian}) is invariant
under $\wil \go \wil +2\pi$ and $p_a\go p_a -1$.   In other words,
\beqn
&&U= \exp \Big( 2\pi i\Pw + i\sum_{a=1}^{N} q_a \Big) \cr
&&[U, H_\tot]=0  \cr
&&U \, |\Phi_\vac (\theta) \ra = e^{i\theta} \,  |\Phi_\vac (\theta) \ra ~~,
\label{GaugeInv1}
\eeqn
which implies the vanishing of 
$\la e^{ik\wil}\ra_{\theta}$  for a non-integer $k$.

In the $N=1$ (one-flavor) case, there is no $\vphi_a$ degrees of freedom.
Eq.\ (\ref{Polyakov2}) reduces to
$\la P_{ke} \ra_{\theta,T} = e^{-k^2\pi \mu /4T}$, which agrees with the 
result of Grignani et al.  The factor $k^2 \pi\mu/4$ is understood
as the self-energy of the source \cite{Grignani1}.  

In the $N\ge 2$ (multi-flavor) case 
the overlap integral for the $f(\vphi;\theta_\eff)$ factor 
becomes relevant.  In the massless ($m=0$) case, however, $f(\vphi)$ is
constant as the potential $\potN$ in (\ref{QMeq}) vanishes.  Hence 
$\la P_{ke} \ra_{\theta,T} = e^{-k^2\pi \mu /4NT}$ in the massless theory.

When $m\not=0$, the overlap integral needs to be evaluated numerically.  In  
two limits, namely $T/\mu \ll (m/\mu)^{N\over N+1}$ and $T/\mu \gg 1$, analytic
expressions are obtained.  It is instructive to examine the free energy
$F_e(T) = - T \ln \la P_{e} \ra_{\theta,T}$ for $k=1$.  At sufficiently low
$T=L^{-1}$,
$\kappa_0 \gg 1$ in (\ref{Potential2}) so that $f(\vphi)$ has a sharp peak
at the minimum of the potential  
\beeq
\vphi_a^{\rm min} = {\bar \theta_\eff\over N}  \next 
\bar\theta_\eff = \theta_\eff - 2\pi \Big[ {\theta_\eff + \pi \over 2\pi} \Big]
\label{minimum1}~~~.
\eneq
$f(\vphi)$ is approximately given, up to a normalization constant,  by
\beeq
f = \exp \Bigg\{ - \sqrt{ {N-1\over 2N} \,
      \kappa_0 \cos {\bar \theta_\eff\over N}} 
~ \bigg(  \sum_{a=1}^{N-1} \tilde\vphi_a^2
+\sum_{a<b}^{N-1} \tilde\vphi_a \tilde\vphi_b \bigg) \Bigg\}
\next \tilde \vphi_a= \vphi_a - {\bar\theta_\eff\over N}~~. 
\label{waveFunction2}
\eneq
Hence the overlap integral gives an additional damping factor in
(\ref{Polyakov2}).   In the opposite limit $T/\mu \gg 1$, $\kappa_0 \ll 1$ so
that
$f \sim$ constant. Hence we find, for an integer $k=q/e$,
\beeq
F_q(T) = \cases{
\myfrac{k^2\pi\mu}{4N}
 \Bigg\{ 1 + (N-1) \bigg( 2 e^\gamma \myfrac{m}{\mu} 
      \cos \myfrac{\bar\theta_\eff}{N} \bigg)^{N\over N+1} \Bigg\}
   &for $T \ll m^{N\over N+1} \mu^{1\over N+1}$\cr
\myfrac{k^2\pi\mu}{4N} &for $T \gg \mu$~~~.\cr}
\label{FreeEnergy1} 
\eneq
The free energy is finite. It does not diverge even at $T=0$.  An integer 
external charge is screened.

%%%%%%%%%%%%%%%%%%%%%%%%  figure 1  %%%%%%%%%%%%%%%%%%%%%%%%%%
\begin{figure}[tb]
\epsfxsize= 10.cm    % 11.cm is too big
%\epsffile[0 120 400 650]{P-loop.ps} % works ok
\epsffile[0 50 400 550]{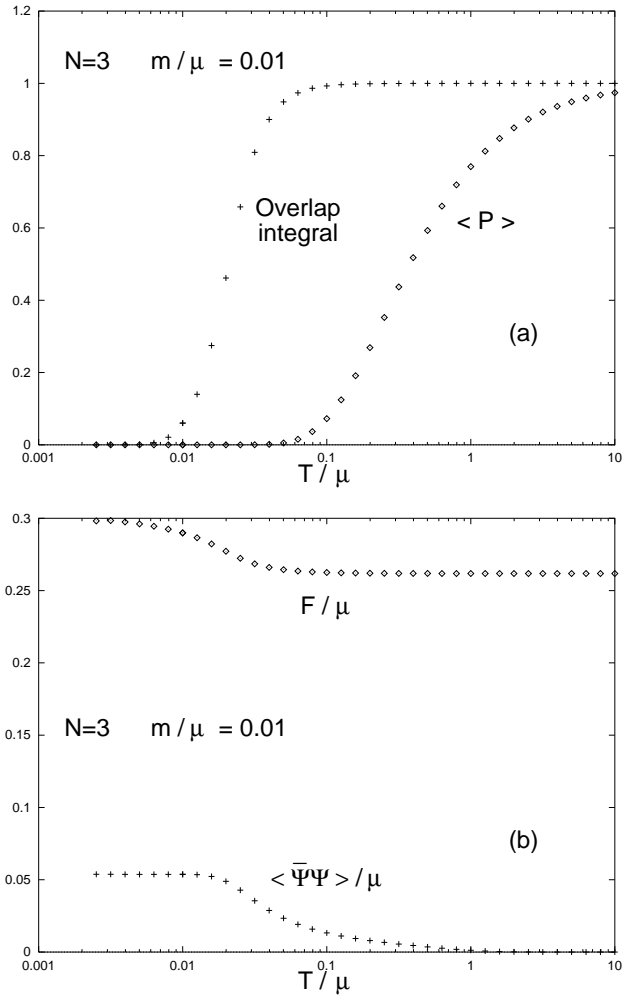}
\vskip -0cm
\caption{(a) The Polyakov loop  $\la P \ra$  and the overlap integral in (16)
%\protect\( \ref{Polyakov2} \protect\) 
are plotted as functions of $T/\mu$ for $k=1$ ($q=e$),
$N=3$, $m/\mu=0.01$ and  $\theta=0$.  (b)  The free energy $F/\mu= -(T/\mu) \ln
\la P \ra$ and the chiral condensate $\la \psibar\psi\ra/\mu$ are plotted with
the same parameter values.}
\label{fig.1}
\vskip 0.3cm
\end{figure}

%%%%%%%%%%%%%%%%%%%%%%%%%%%%%%%%%%%%%%%%%%%%%%%%%%%%%%%%%%%%%%

In fig.\ 1 we have depicted the temperature dependence of the Polyakov loop,
free energy, and chiral condensate in the $N=3$ case with $m/\mu=0.01$ and
$\theta_\eff=0$.  Notice that all these quantities show cross-over transitions,
but at different temperature.

The vanishing of the Polyakov loop for a fractional charge $q=ke$ 
does not necessarily implies the confinement as it follows from the gauge
invariance. To obtain information on the confinement or the string tension, we
evaluate the Polyakov loop correlator:
\beeq
\tilde G_q(x) = \la P_q(x)^\dagger P_q(0) \ra_{\theta,T}
\Longleftrightarrow
G_{q=ke}(t) = \la T[\, e^{-ik\wil(t)}\, e^{ik\wil(0)}\, ]\ra_{\theta,L}
 ~~.
\label{correlator}
\eneq
Without loss of generality we suppose that $x>0$ and $t>0$ ($x \leftrightarrow
it$).  Note that $G_q(t)$ is gauge invariant.

In the $N=1$ model the correlator $G_e(t)$ was first evaluated
by Hetrick and Hosotani \cite{HH}.  
For a general $q$,  $\tilde G_q(x)$ has been
evaluated by Hansson et al. \cite{HNZ}
and by Grignani et al. \cite{Grignani1}. 
Consider
\beeq
G(t; k,l) = \la T[\,e^{-ik\wil(t)} \, e^{+il\wil(0)} 
\, ]\ra_{\theta,L}
\label{correlator2}
\eneq
in the massless $N$-flavor model, which we denote by $G^{(0)}(t;k,l)$.
In this case the zero modes ($\wil, q_a$) and oscillatory
modes decouple so that the model is exactly solvable.  The Heisenberg
operator $\wil(t)$ can be expressed in terms of Schr\"odinger operators
\beeq
\wil(t) = \wil \cos\mu t + {\pi\mu L\over N} \, \Pw \sin \mu t
+ {2\pi\over N} \sum_{a=1}^N p_a (\cos \mu t - 1) ~.
\label{Hoperator}
\eneq

Making use of (\ref{Hoperator}), we find 
\beqn
&&e^{-ik\wil(t)}\, e^{+il\wil(0)} =
\exp \bigg\{ -{ikl\pi\mu L\over 2N} \, \sin \mu t 
+ {ik\pi\mu L\over 2N} \, \sin\mu t \, (k\cos\mu t - l) \bigg\}\cr
\noalign{\kern 10pt}
&&\hskip 1cm \times 
\exp \bigg\{ -{2ik\pi\over N}  (\cos\mu t -1)  \sum p_a
-i(k\cos\mu t - l) \wil \bigg\} \cr
\noalign{\kern 10pt}
&&\hskip 1cm \times
\exp \bigg\{ - {ik\pi\mu L\over N} \, \sin \mu t \, \Pw \bigg\}~~~.
\label{identity1}
\eeqn
In taking the vacuum expectation value of (\ref{identity1}), we 
encounter the factor $\delta_{2\pi}[\theta-\theta'+2\pi(k-l)]$
as in (\ref{Wilson2}).   It follows that
\beeq
G^{(0)}(t;k,l) = \cases{
0 &for $k-l\not=$ an integer\cr
\exp \bigg\{ - {\mybig\pi\mu L\over\mybig 4N} \,
(k^2+l^2-2kl e^{-i\mu|t|} ) \bigg\} &for $k-l=$ an integer.\cr}
\label{correlator3}
\eneq
This implies that the increase in the free energy in the presence
of a pair of charges $q$ and $-q$ is
\beeq
F_q^{\rm pair}(T)^{(0)} = - T \ln \la P_q(x)^\dagger P_q(0) \ra_\theta
= {\pi\mu\over 2N} \Big({q\over e}\Big)^2 \, (1- e^{-\mu|x|}) ~.
\label{FreeEnergy2}
\eneq
In the massless theory external charges are completely shielded and
the string tension vanishes as shown by various authors
\cite{CJS,HNZ,Gross,Grignani1}.

If fermions are massive, the situation qualitatively changes.
In the literature only the $N=1$ case has been analysed for which 
a perturbation theory in fermion masses is valid.  Restricting
ourselves to $N=1$ with $k=l$ and $\delta=0$ (therefore $\theta_\eff=\theta$),
we find
\beqn
G^{(1)}(t;k,k) &=& G(t;k,k) - G^{(0)}(t;k,k)  \cr
\noalign{\kern 6pt}
\hskip .2cm &=& -i \int_{-\infty}^\infty ds \, \Big\{ 
\la T[\, e^{-ik\wil(t)^I} \, e^{il\wil(0)^I} \, H_\rint(s)^I ]\ra
- G^{(0)}(t;k,k)  \, \la  H_\rint(s)^I \ra \Big\} \cr
\noalign{\kern 12pt}
H_\rint(s)^I &=& -{m B(\mu L)\over L} \int_0^L dx\,
 \Big\{ e^{iq(s)^I} N_\mu[e^{i\sqrt{4\pi} \phi(s,x)^I}] + h.c. \Big\}~~.
\label{correlator4}
\eeqn
Here the superscript $I$ indicates the interaction picture defined by a
massless fermion.  To O($m$) the $\phi$ field part of $H_\rint$ does not
contribute.

In the second term in the expression of $G^{(1)}$, 
$\la H_\rint (s) \ra = mB e^{-\pi/\mu L} \cos\theta$.  In evaluating the first
term we need, in addition to (\ref{Hoperator}), 
\beeq
q(t)^I= q + {2\over \mu L}  \, (\wil + 2\pi p)  \sin \mu t 
+ 2\pi \Pw  (1-\cos \mu t) ~~.
\label{Hoperator2}
\eneq
A useful identity is
\beeq
\la e^{\pm i q(s)^I} \, e^{-ik\wil(t)^I} \, e^{ik\wil(0)^I} \ra_\theta
= G^{(0)}(t;k,k)
\cdot e^{\mp i\theta} \, e^{-\pi/\mu L} \,
     e^{\mp i\pi k (1 - e^{i\mu t}) e^{-i\mu s}} ~~~.
\label{identity2}
\eneq

Without loss of generality we take $t>0$.   The integral
over $s$ in (\ref{correlator4})  splits into three parts: $\int_{-\infty}^0$,
$\int_0^t$, and $\int_t^\infty$.  It is easy to check that the first integral
is the same as the  third integral after a change of variables, and each of
them vanishes.  The manipulation is justified with the hypothesis of
adiabatic switching of interactions implicit in the derivation of 
Gell-Mann-Low relations.

The second integral gives the sole contribution to $G^{(1)}$:
\beqn
G^{(1)}(t;k,k) = imB e^{-\pi/\mu L} G^{(0)}(t;k,k) \hskip 6cm &&\cr
\noalign{\kern 10pt}
\times \int_0^t ds \bigg\{ e^{-i\theta} \Big( 
e^{2\pi ik} e^{-i\pi k ( e^{-i\mu (t-s)} + e^{-i\mu s}) } - 1 \Big)
 + (\theta \go - \theta, k \go -k) \bigg\}~~. &&
\label{correlator5}
\eeqn
The integral is expressed in terms of Bessel functions.
The correction to the free energy is, after making a Wick rotation $it=x>0$,
\beqn
F_q^{\rm pair} (x,T)^{(1)} &=&  -T \, G^{(1)} \Big/  G^{(0)}\cr
\noalign{\kern 10pt}
&=& - 2|x| mTB e^{-\pi/\mu L} \Bigg\{
 J_0(2\pi kz) \cos(\theta-2\pi k) - \cos \theta   \cr
\noalign{\kern 10pt}
&& +{1\over \mu|x|} 
 \sum_{n=1}^\infty \bigg(  e^{-i(\theta-2\pi k)} 
        + (-1)^n e^{+i(\theta-2\pi k)} \bigg) \, {i^n\over n} 
(z^{-n} - z^n) J_n(2\pi kz)  \Bigg\} 
\label{FreeEnergy3}
\eeqn
where $z=e^{-\mu|x|/2}$.
For $\mu|x| \gg 1$, $z\ll 1$ so that
\beqn
&&F_q^{\rm pair} (x,T)^{(1)} \sim \sigma \, |x| \cr
\noalign{\kern 10pt}
&&\sigma^{N=1} = - 2mT B\Big( {\mu\over T} \Big) \, e^{-\pi T/\mu} \,
   \Big\{ \cos(\theta-{2\pi q\over e}) - \cos \theta \Big\}  ~. 
\label{StringTension1}
\eeqn
Here $\sigma$ is a ``string tension''.  Since $\la \psibar\psi\ra_\theta
= - 2T e^{-\pi T/\mu} B(\mu/T) \cos\theta$, we find
\beeq
\sigma^{N=1} = m \Big\{ \la \psibar \psi \ra_{\theta-2\pi (q/e)} - \la
\psibar\psi\ra_\theta \Big\}  ~~~.
\label{StringTension2}
\eneq
In other words, the major effect of a pair of external sources of charges
$q$ and $-q$ is to shift the $\theta$ parameter in the region bounded by the
sources by an amount $2\pi(q/e)$, which changes the chiral condensate
\cite{CJS,HNZ}.  A linear potential results because of this.  We shall show 
below that this is true even for $N\ge 2$.   The expression
(\ref{StringTension2}) is valid at arbitrary  temperature. 

The string tension $\sigma$ can be either positive or negative, depending on the
values of  $\theta$ and $q/e$.  This implies that the $\theta\not=0$ vacuum
is unstable against pair creation  of sufficiently small fractional charges.

The perturbation theory in fermion masses cannot be employed in the $N\ge 2$
case as physical quantities are not analytic in $m$ at $T=0$
\cite{Coleman,HHI}.  The  perturbation theory can be applied only in the high
temperature regime.

There is a better way to explore the problem.  We place external charges
on a circle and solve the Hamiltonian as was done
 in the $N=1$ case in \cite{HH} and \cite{Iso}.  

In the presence of external charges $\L_\ext= - A_0 \rho_\ext$.  
Gauss's law implies
\beeq
\dd_x E_\ext(x) = \rho_\ext(x) ~~~.
\label{Gauss}
\eneq
Let us restrict ourselves to  static sources $\rho_\ext(x)$ where
 $\int_0^L dx \, \rho_\ext =0$.  Then 
$E_\ext(x) = E_\ext^{(0)}- A_0^\ext(x)'$.  Here $E_\ext^{(0)}$ is constant and
$A_0^\ext(x) = - \int_0^L dy\, G(x-y) \rho_\ext(y)$.  In particular, for a pair
of sources located at $x=0$ and at $x=d$,
\beqn
\rho_\ext(x) &=& q \Big\{ \delta_L(x) - \delta_L(x-d) \Big\} \cr
\noalign{\kern 12pt}
E_\ext(x) &=& E_\ext^{(0)} + E_\ext(x)^{(1)} \cr
\noalign{\kern 12pt}
E_\ext^{(1)} &=& - A_0^\ext(x)' = \cases{
q \Big( 1 - \myfrac{d}{L} \Big) &for $0<x<d$\cr
-q \myfrac{d}{L} &for $d<x<L$~~~.\cr}
\label{externalE}
\eeqn
Note that $\int_0^L dx \, E_\ext(x)^{(1)}=0$.
% We argue that $E_\ext^{(0)}= qd/L$ when $d \ll L$ so that $E_\ext(x)\sim 0$
% for $d<x<L$.  

Suppose that $m_a=m\ll \mu$.
The total charge density is $j^0_\EM = \sum_{a=1}^N e \psi_a^\dagger
\psi_a + \rho_\ext$, and   the Coulomb energy becomes
\beqn
H_{\rm Coulomb} &=&
- {1\over 2} \int_0^L dxdy \, j^0_\EM (t,x) G(x-y)  j^0_\EM (t,y) \cr
\noalign{\kern 10pt}
&=& \int_0^L dx \, {1\over 2} (\mu \chi_1 - E_\ext^{(1)})^2~~~.
\label{Coulomb1}
\eeqn
Here $\chi_1 = N^{-1/2} \sum_{a=1}^N \phi_a$.  In view of  (\ref{Coulomb1}) we
write
\beqn
&&\chi_1 = \chi_1^{cl} + \tchi_1 \cr
\noalign{\kern 10pt}
&&\Big( - {d^2\over dx^2} + \mu^2 \Big) \, \chi_1^{cl} = \mu\, E_\ext^{(1)}
~~.
\label{chi-classical}
\eeqn
The total Hamiltonian is now
\beqn
&&H_\tot^{\rm new} = H_0 + H_{cl} + H_\chi + H_\mass\cr
\noalign{\kern 10pt}
&&H_{cl} = \int_0^L dx {1\over 2} \bigg\{
(\chi_1^{cl}{\,}')^2 + (\mu \chi_1^{cl} - E_\ext^{(1)})^2 \bigg\} \cr
\noalign{\kern 10pt}
&&H_\chi = \int_0^L dx {1\over 2} :\, \Bigg\{
 \Pi_1^2 + (\tchi_1')^2 + \mu^2 \tchi_1^2 
+ \sum_{\alpha=2}^N (\Pi_\alpha^2 + \chi_\alpha'^2) \Bigg\} :\,
\label{newHamiltonian}
\eeqn
$H_0$ and $H_\mass$ are given by (\ref{Hamiltonian}).   When $m\ll \mu$,
$\mu_1 \sim \mu$.   In $H_\mass$
\beeq
N_{\mu}[e^{i \sqrt{4\pi/N} \chi_1}]
= e^{i\sqrt{4\pi/N} \chi_1^{cl}} \, N_{\mu}[e^{i \sqrt{4\pi/N} \tchi_1}] ~~.
\label{newMass}
\eneq
In other words, the net effect of a pair of external sources is to give
 $x$-dependent fermion mass phases $\delta_a= 
\sqrt{4\pi/N} \chi_1^{cl}(x)$ in (\ref{Lagrangian}).
Suppose that $\mu^{-1} \ll d\ll L$.  Sufficiently away from the source
\beqn
\delta_a^\eff &=& 
\sqrt{{4\pi\over N}} \, \chi_1^{cl}(x) \sim {2\pi\over Ne} \, E_\ext(x)^{(1)}\cr
\noalign{\kern 10pt}
&=& \cases{  \myfrac{2\pi q}{Ne} \Big(1-\myfrac{d}{L} \Big)
 \equiv \delta^\in  &for $\mu^{-1} < x < d-\mu^{-1}$\cr
 - \myfrac{2\pi q}{Ne} \cdot \myfrac{d}{L} \equiv \delta^\out 
  &for $d+\mu^{-1} < x < L-\mu^{-1}$~~~.\cr}
\label{externalE2}
\eeqn

Finding the exact form of the  ground state wave function of
(\ref{newHamiltonian}) is rather involved.   Instead, we content ourselves
with finding an approximate wave function, noticing that $\delta_a^\eff$ is
almost constant between the two sources.  

The entire circle  is divided into the two regions,   the inside region
 $0<x<d$ and outside region $d<x<L$ ($\mu^{-1} \ll d\ll L$).  
For the evaluation of local physical quantities in each region, one can
approximately write the ground state as a direct product of ground states 
in the two regions:  $|\Psi_g\ran \sim |\Psi\ran_\in \otimes
|\Psi\ran_\out$.  In the absence of sources,  $|\Psi_g\ran \sim |\theta\ran_\in
\otimes |\theta\ran_\out$.  In the presence of sources
\beeq
|\Psi_g \ran 
\sim |\theta +\delta\theta ; \delta^\in \ran_\in \otimes 
       |\theta+\delta\theta; \delta^\out \ran_\out  ~~~.
\label{newGroundState}
\eneq
In addition to the effect of $\delta_a^\eff$ an overall shift $\delta\theta$ in
the  $\theta$ value results as $\delta_a^\eff$ is $x$-dependent.   After all
there is only one $\theta$ parameter globally.

To determine $\delta\theta$, we utilize the fact that local physical
quantities in the infinite volume limit $L\go\infty$ must reproduce
results in the Minkowski spacetime.  In particular, physics in the
outside region
$d+\mu^{-1} < x  < L- \mu^{-1}$ ($L\go\infty$) must be essentially the same
as physics in the absence of sources.  In other words
$|\theta+\delta\theta; \delta^\out \ran_\out \sim |\theta; \delta^\out=0
\ran_\out$. As shown above physics depends on $\theta$ through the combination
$\theta_\eff=\theta- \sum_{a=1}^N \delta_a$. This determines
\beeq
\delta\theta = N \delta^\out = -{2\pi q\over e} \cdot {d\over L}
\label{dtheta}
~~~.\eneq
Hence, the net effect is summarized by
\beqn
&&|\Psi_g \ran 
\sim |\theta_\eff \ran_\in \otimes 
       |\theta  \ran_\out \cr
\noalign{\kern 10pt}
&&\theta_\eff = \theta - {2\pi q\over e}  ~~~.
\label{newGroundState2}
\eeqn
Consequently, the change in the energy due to the external sources is,
to O($d/L$), 
\beeq
\Delta E =  N m d \Big\{ \la \psibar\psi \ra_{\theta_\eff}
                         - \la \psibar\psi \ra_{\theta} \Big\}
\eneq
so that the string tension is 
\beeq
\sigma = N m \Big\{ \la \psibar \psi \ra_{\theta-2\pi (q/e)} - \la
\psibar\psi\ra_\theta \Big\}  ~~~.
\label{StringTension3}
\eneq
The result generalizes to finite temperature, as  
was seen above in the $N=1$ case.

Note that the parameter $\theta$ is not completely equivalent to the 
electric field $E$.  Indeed, in the absence of sources 
$\la E\ra_\theta = \la e P_W - A_0' \ra_\theta =0$.  
External sources, or external electric fields, induce effective fermion mass
phases $\delta_a^\eff$ in the Hamiltonian, which in turn changes the 
effective $\theta_\eff$ through the chiral anomaly.   We also remark that the
Coulomb energy (\ref{Coulomb1}) is O[$(d/L)^0$].  The linear potential
results from the change in the chiral condensate.

The chiral condensate $\la \psibar \psi\ra_\theta$ at arbitrary temperatue
$T=L^{-1}$ has been evaluated in ref. \cite{HHI}.  With given $m$, the dependence
of $\sigma$ on charge $q$ is essentially the $\theta$ dependence of
$\la \psibar \psi\ra_\theta$.  At $T=0$ it has a cusp at $\theta_\eff=\pi$
($mod ~ 2\pi$).  More explicitly
\beeq
\sigma_{\theta}^{T=0} = -  \mu^2{N \over 2\pi}
\bigg( 2 e^\gamma\, {m\over \mu} \bigg)^{{2N\over N+1}}
\Bigg\{ \bigg( \cos {\bar\theta_\eff\over N} \bigg)^{{2N\over N+1}}
 - \bigg( \cos {\bar\theta\over N} \bigg)^{{2N\over N+1}} \Bigg\} 
 ~~.
\label{StringTension4}
\eneq
Notice the singular dependence of $\sigma$ on $m$ as well.  A mass perturbation 
theory cannot be employed at low temperature for $N\ge 2$.

In the high temperature limit
\beeq
 \la \psibar\psi \ra_\theta 
   = - {2N\over \pi(N-1)} \, m ~
\cases{
 \bigg( {\mybig \mu e^\gamma \over\mybig 4\pi T} \bigg)^{2/N}  
  &for $m^{N\over N+1} \mu^{1\over N+1} \ll T \ll \mu$\cr\cr
  e^{-2\pi T/N\mu} &for $T \gg \mu$\cr} 
  \label{condensateN-highT}
\eneq
for $N\ge 3$.  There appears no $\theta$ dependence to this order.  Hence
the string tension $\sigma$ is at most O($m^3$) in this regime.

For $N=2$, the expressions for  $\la \psibar\psi \ra_\theta$   in
(\ref{condensateN-highT}) is multiplied by a factor
 $2 \cos^2 \onehalf\theta $.  Therefore
\beeq
\sigma^{N=2}_\theta = - {4\over \pi} m^2 
(\cos \theta_\eff - \cos\theta )
\times \cases{
 {\mybig \mu e^\gamma \over\mybig 4\pi T} 
  &for $m^{2\over 3} \mu^{1\over 3} \ll T \ll \mu$\cr\cr
  e^{-\pi T/\mu} &for $T \gg \mu$~~.\cr} 
  \label{StringTension5}
\eneq

%%%%%%%%%%%%%%%%%%%%%%%%%%%%%%  figure 2  %%%%%%%%%%%%%%%%%%%%%%%%%%%
\begin{figure}[tb]
\epsfxsize= 13.0cm    % changed from 10 cm  to 8.5 cm.  9cm is too big.
%\epsffile[100 40 650 500]{S-tension.ps}
\epsffile[0 40 450 300]{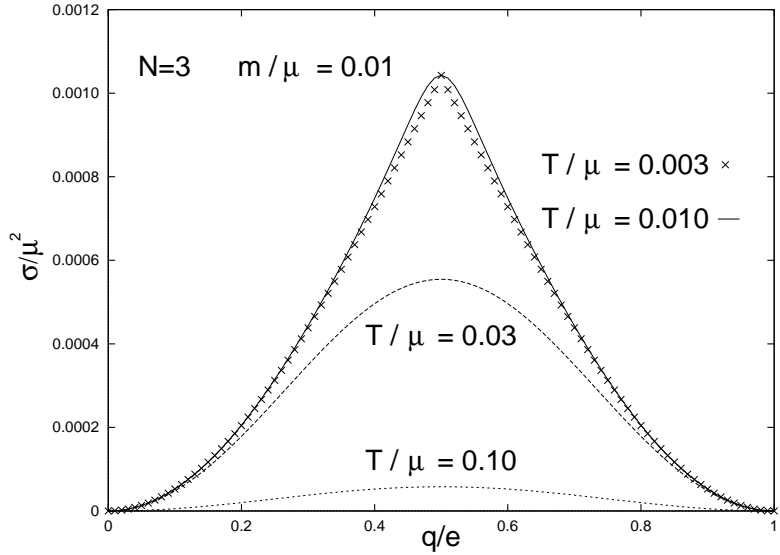}
%\epsffile[120 120 470 650]{S-tension.ps}   %---
\vskip -0cm
\caption{String tension $\sigma/\mu^2$ in (45) is depicted as a function of
$q/e$ at various temperature in the $N=3$ case with $m/\mu=0.01$ and 
$\theta=0$ .  A cusp develops at
$q/e=0.5$ in the 
$T\go 0$ limit.}
\label{fig:2}
\vskip 0.3cm
\end{figure}
%%%%%%%%%%%%%%%%%%%%%%%%%%%%%%%%%%%%%%%%%%%%%%%%%%%%%%%%%%%%%%%%%%

For a general value of $T/\mu$, $\sigma$ must be evaluated  numerically.
In fig.\ 2 we have displayed  $q/e$ dependence of $\sigma/\mu^2$ at
$T/\mu=0.003$, 0.01, 0.03, and 0.1 with $m/\mu=0.01$ and $\theta=0$ in the $N=3$
case. One can see how a cusp behavior develops at $q/e=0.5$ ($\theta_\eff=\pi$)
as the temperature goes down.  At higher temperature the magnitude of the 
string tension rapidly diminishes.  For instance,
  $\sigma/\mu^2=1.8 \times 10^{-7}$ at $T/\mu=1$ and $q/e=0.5$ .

In this paper we have shown that the confinement of fractional charges in the 
massive $N$-flavor Schwinger model results from the effective change in
the $\theta$ parameter which alters chiral condensates.  In the multi-flavor
case ($N\ge 2$) the string tension at zero temperature has singular dependence
on fermion masses and the $\theta$-parameter.

\vskip 1cm
\leftline{\bf Acknowledgments}

This work was supported in part by by the U.S.\ Department of Energy
under contracts DE-FG02-87ER-40328 (R.R.) and by 
DE-AC02-83ER-40105 (Y.H.).   Y.H.\ would like to thank Jim Hetrick and 
Satoshi Iso for useful communications.

%\eject
%\myend

\vskip .5cm 

\def\ap {{\it Ann.\ Phys.\ (N.Y.)} }
\def\cmp {{\it Comm.\ Math.\ Phys.} } 
\def\ijmpA {{\it Int.\ J.\ Mod.\ Phys.} {\bf A}} 
\def\ijmpB {{\it Int.\ J.\ Mod.\ Phys.} {\bf B}} 
\def\ijmpC {{\it Int.\ J.\ Mod.\ Phys.} {\bf C}} 
\def\jmp {{\it  J.\ Math.\ Phys.} } 
\def\mplA {{\it Mod.\ Phys.\ Lett.} {\bf A}} 
\def\mplB {{\it Mod.\ Phys.\ Lett.} {\bf B}} 
\def\plB {{\it Phys.\ Lett.} {\bf B}} 
\def\plA {{\it Phys.\ Lett.} {\bf A}} 
\def\nc {{\it Nuovo Cimento} } 
\def\npB {{\it Nucl.\ Phys.} {\bf B}} 
\def\pr {{\it Phys.\ Rev.} } 
\def\prl {{\it Phys.\ Rev.\ Lett.} } 
\def\prB {{\it Phys.\ Rev.} {\bf B}} 
\def\prD {{\it Phys.\ Rev.} {\bf D}} 
\def\prp {{\it Phys.\ Report} } 
\def\ptp {{\it Prog.\ Theoret.\ Phys.} } 
\def\rmp {{\it Rev.\ Mod.\ Phys.} } 

\myend

%%%% comment %%%%
\begin{thebibliography}{99}

\baselineskip=14pt
\parskip=0pt

\small

%Seminal papers
\bibitem{SchwGEN}
 J. Schwinger, \pr {\bf 125} (1962) 397 ;  {\bf 128} (1962) 2425 .
\bibitem{Lowenstein1} 
J.H. Lowenstein and J.A. Swieca, \ap {\bf 68} (1971) 172 .
\bibitem{Casher} 
A. Casher, J. Kogut and L. Susskind, \prl {\bf 31} (1973) 792 ; 
\prD {\bf 10} (1974) 732 .
\bibitem{CJS}
S. Coleman, R. Jackiw, and L. Susskind,  \ap {\bf 93} (1975) 267 .
\bibitem{Coleman} S. Coleman,  \ap {\bf 101} (1976) 239 .
\bibitem{Nielsen} N.K. Nielsen and B. Schroer, \npB {\bf 120} (1977) 62 .

%  more
\bibitem{Morchio} G. Morchio, D. Pierotti and F. Strocchi,
    \ap {\bf 188} (1988) 217.  % confinement
\bibitem{Ellis} J.\ Ellis, Y.\ Frishman, A.\ Hanany, M.\ Karliner, 
       \npB {\bf 382} (1992) 189 .  % soliton


\bibitem{Smilga} A.V. Smilga, \plB{\bf 278} (1992) 371 ;
\prD {\bf 46} (1992) 5598 ; \prD {\bf 49} (1994) 5480 ; 
\ap {\bf 234} (1994) 1 .

\bibitem{Seiler} C. Gattringer and E. Seiler, \ap {\bf 233} (1994) 97 .
   %  functional integration
\bibitem{Gattringer1} C.\ Gattringer, {\tt hep-th/9503137, 
    hep-th/950592}, MPI-PH-95-52.

% recent progress

% Condensates, confinement and screening
\bibitem{HNZ}  T.H. Hansson, H.B. Nielsen and I. Zahed,
        \npB {\bf 451} (1995) 162 .
\bibitem{HHI}  J.E.\ Hetrick, Y.\ Hosotani and S.\ Iso, \plB {\bf 350} 
(1995) 92 , {\tt hep-th/9510090} (to appear in \prD); 
Y.\ Hosotani {\tt hep-ph/9510387}. 
\bibitem{Gross} D.J.\ Gross, I.R.\ Klebanov, A.V.\ Matytsin, A.V.\ Smilga, 
        {\tt hep-th/9511104}.
\bibitem{Grignani1} G.\ Grignani, G.\ Semenoff, P.\ Sodano, O.\ Tirkkonen, 
      {\tt hep-th/9511110}.
\bibitem{Smilga2} A. Smilga and J.J.M. Verbaarschot,  {\tt hep-ph/9511471}.

%  massless, N flavor
\bibitem{Halpern} M.B. Halpern, \prD {\bf 13} (1976) 337 . 
\bibitem{Affleck} I. Affleck, \npB {\bf 265} [FS15] (1986) 448 .



% Poliakov Loop
\bibitem{Polyakov} A.M.\ Polyakov, \plB {\bf 72} (1978) 477 .
\bibitem{Susskind} L.\ Susskind, \prD {\bf 20} (1979) 2610 .

   %  N=1  effective action,  confinement
\bibitem{Actor} A.\ Actor, \ap {\bf 159} (1984) 445 .  

\bibitem{Svetitsky} B.\ Svetitsky, \prp {\bf 132} (1986) 1 .
\bibitem{Grignani2} G.\ Grignani, G.\ Semenoff, P.\ Sodano, 
{\tt hep-th/9503109, hep-th/9504105}

% S^1,  Wison line correlation, external charge
\bibitem{HH} J.E. Hetrick and Y. Hosotani, \prD {\bf 38} (1988) 2621 .

\bibitem{Iso} S. Iso and H. Murayama, \ptp {\bf 84} (1990) 142 .


% Circle 
%%%%% Massless
%%%%%%%%%%%%%% N=1 
\bibitem{Wolf}  D. Wolf and J. Zittartz, {\it Z. Phys.} {\bf B59} (1985) 117.
\bibitem{Manton}  N. Manton, \ap {\bf 159} (1985) 220 .
\bibitem{SchwS1} R. Link, \prD {\bf 42} (1990) 2103 .
\bibitem{Sara} F.M. Saradzhev, \plB{\bf 278} (1992) 449 ; 
   \ijmpA{\bf 9} (1994) 3179, {\tt hep-th/9501001}.

%   Topological SM
\bibitem{Itoi} C. Itoi and H. Mukaida, \mplA{\bf 7} (1992) 259 .
%---------------------  (Instanton) ----------------------------------
\bibitem{Ross} M.B. Paranjape and R. Ross, \prD {\bf 48} (1993) 3891 .
\bibitem{Paranjape} M.B. Paranjape,  \prD {\bf 48} (1993) 4946 .

\bibitem{Shifman} M.A. Shifman and A.V. Smilga, \prD {\bf 50} (1994) 7659 .
%-------------------- (Hamiltonian Reduction) ------------------------
\bibitem{Itakura} K. Itakura and K. Ohta, \prD {\bf 50} (1994) 4145 .
%--------------------- (Two body rel.) ------------------------------
\bibitem{Barut} A.O. Barut and F.M. Saradzhev, \ap {\bf 235} (1994) 220 .
        % 2 body QM


%--------------------- ( Other Topology )  ---------------------------

% finite tempeature
\bibitem{Love} S.\ Love, \prD {\bf 23} (1981) 420 . %thermodynamics
\bibitem{Stam} K.\ Stam and J.\ Visser, J.\ Phys.\ {\bf G11} (1985) L:143 .
\bibitem{Ruiz} F.\ Ruiz\ Ruiz and R.F.\ Alvarez-Estrada, 
\plB {\bf 180} (1986) 153 .
\bibitem{Das} A.\ Das and A. Karev, \prD {\bf 36} (1987) 623 .
\bibitem{Thom} G.\ Thompson and R.B.\ Zhang, J.\ Phys.\ {\bf G13} 
(1987) L:93 .  
\bibitem{Baier} R.\ Baier and E.\ Pilon, Z.\ Phys.\ {\bf C52} (1991) 339 .

\bibitem{SchwFT} I.\ Sachs and A. Wipf, {\it Helv. Phys. Acta.} 
{\bf 65} (1992) 652 .
%   Chiral condensate
\bibitem{Kao} Y.C.\ Kao, Mod.\ Phys.\ Lett.\ {\bf A7} (1992) 1411 .
\bibitem{Fay} A.\ Fayyazuddin, T.H.\ Hansson, M.A.\ Nowak, J.J.M.\ 
Verbaarschot and I.\ Zahed, \npB {\bf 425} (1994) 553 .  % correlator

\bibitem{Kiefer} C. Kiefer and A. Wipf, \ap {\bf 236} (1994) 241 .
\bibitem{Steele} J.\ Steele, A.\ Subramanian and I.\ Zahed, 
\npB {\bf 452} (1995) 545 .  % correlator



%Torus
\bibitem{Joos1} H. Joos, {\it Helv. Phys. Acta.} {\bf 63} (1990) 670 .
\bibitem{Joos2} H. Joos and S.I. Azakov, {\it Helv. Phys. Acta.} {\bf 67}
(1994) 723 .
\bibitem{Joos3} H. Joos,  {\it The non-perturbative structure of gauge
   theories with massless fermions}   (Dec 1995).



%S2
\bibitem{Jaye} C. Jayewardena, {\it Helv. Phys. Acta.} {\bf 61} (1988) 636 .
%Bag
\bibitem{Wipf} A. Wipf and S. D\"urr, ZU-TH 30/94, ETH-TH/94-36, 
\npB{\bf 443} (1995) 201.
%--------------------- (on Riemann Surfaces)---------------------------
\bibitem{Ferrari} F. Ferrari,  {\tt hep-th/9310024};
      {\tt hep-th/9310055}, \npB{\bf 439} (1995) 692 .   
%----------------------------------------------------------------------


% lattice
%--------------------- (Light cone) ----------------------------------
\bibitem{Lenz} F. Lenz, M. Thies, S. Levit and K. Yazaki,
\ap {\bf 208} (1991) 1 .
\bibitem{Heinzl} T. Heinzl, S. Krusche, and E. Werner, 
\plB{\bf 275} (1992) 410 .

\bibitem{Yee} K. Yee, \prD {\bf 47} (1993) 1719
\bibitem{Dilger} H. Dilger, (Dec 93)  DESY 93-181 ;
     \npB{\bf 434} (1995) 321,  \ijmpC {\bf 6} (1995) 123. 
\bibitem{Dilger2} H. Dilger and H. Joos, \npB (Proc. Suppl.) {\bf 34}
  (1994) 195.
\bibitem{Harada} K. Harada, T. Sugiura and M. Taniguchi,
   \prD {\bf 49} (1994) 4226.
\bibitem{Belyaev}  V.M. Belyaev, {\tt hep-lat/9412033}
\bibitem{Irving} A.C. Irving and J.C. Sexton, {\tt hep-lat/9508032}
\bibitem{Horvath} I. Horvath, {\tt hep-lat/9510018}.

\bibitem{Hallin} J. Hallin and P. Liljenberg, {\tt hep-th/9601136}



\end{thebibliography}
